\documentclass[conference]{IEEEtran}
\usepackage{cite}
\usepackage{graphicx}
\usepackage{amsmath}
\usepackage{times}
\usepackage{latexsym}
\usepackage{graphicx}
\usepackage{bm}
\usepackage{amssymb}
\usepackage[center]{caption2}
\usepackage{stfloats}
\usepackage{cases}
\usepackage{array}
\usepackage{setspace}
\usepackage{fancyhdr}
\usepackage{amscd}
\usepackage{enumerate}
\usepackage{multirow}
\usepackage{float}
\usepackage{color}
\usepackage{longtable}
\usepackage{algorithm}
\usepackage{soul}
\usepackage[caption=false]{subfig}

\usepackage{enumitem}
\usepackage[noend]{algorithmic}
\allowdisplaybreaks

\IEEEoverridecommandlockouts
\begin{document}
\thispagestyle{empty}
\title{Leveraging Synergy of 5G SDWN and Multi-Layer Resource Management for Network Optimization}

\author{
    \IEEEauthorblockN{ Mahsa Derakhshani\IEEEauthorrefmark{2}, Saeedeh Parsaeefard\IEEEauthorrefmark{1}, Tho Le-Ngoc\IEEEauthorrefmark{1}, Alberto Leon-Garcia\IEEEauthorrefmark{2}}
     \IEEEauthorblockA{\IEEEauthorrefmark{2}Department of Electrical \& Computer Engineering, University of Toronto, Toronto, ON, Canada}
    \IEEEauthorblockA{\IEEEauthorrefmark{1}Department of Electrical \& Computer Engineering, McGill University, Montreal, QC, Canada} \\
    \IEEEauthorblockA{Email: mahsa.derakhshani@utoronto.ca; saeideh.parsaeifard@mcgill.ca; tho.le-ngoc@mcgill.ca; alberto.leongarcia@utoronto.ca}
    }
\maketitle

\begin{abstract} 
Fifth-generation (5G) cellular wireless networks are envisioned to predispose service-oriented, flexible, and spectrum/energy-efficient edge-to-core infrastructure, aiming to offer diverse smart-X (city, grid, and phones) applications. Convergence of software-defined networking (SDN), software-defined radio (SDR) compatible with multiple radio access technologies (RATs), and virtualization on the concept of 5G software-defined wireless networking (5G-SDWN) is a promising approach to provide such a dynamic network. The principal technique behind the 5G-SDWN framework is the separation of the control and data planes, from the deep core entities to edge wireless access points (APs). This separation allows the abstraction of resources as transmission parameters of each user over the 5G-SDWN. Similar to traditional wireless networks, in this user-centric, service-oriented and integrated environment, resource management plays a critical role to achieve efficiency and reliability. However, it is natural to wonder if 5G-SDWN can be leveraged to enable converged multi-layer (CML) resource management over the portfolio of resources, and reciprocally, if CML resource management can effectively provide performance enhancement and reliability for 5G-SDWN. We believe that replying to these questions and investigating this mutual synergy are not trivial, but multidimensional and complex for 5G-SDWN, which consists of different technologies and also inherits legacy generations of wireless networks. In this paper, we propose a flexible protocol structure based on three mentioned pillars for 5G-SDWN, which can handle all the required functionalities in a more cross-layer manner compared to the legacy wireless networks (2G, 3G and 4G). Based on this, we demonstrate how the general framework of CML resource management can control the end-user quality of experience. For two scenarios of 5G-SDWN, including both macro-cells and small-cells, we investigate the effects of joint user-association and resource allocation via CML resource management to improve performance in a virtualized network. 

\end{abstract}
%
\section{Introduction}
Cellular wireless networks have been waiting to evolve by 2020 in the context of 5G wireless networks \cite{7113228,7108393}. 5G must offer more service-oriented, flexible, and spectrum/energy-efficient structure to improve quality-of-service (QoS) for end-users by enabling heterogeneity over the utilization of diverse technologies. Besides, with the aim of providing seamless connection for the end-users, all the legacy wireless networks will be integrated under the 5G structure from edge to the core \cite{7113226,7120046,peng2015system}. Among all the new context, three emerging trends in computer and communication networks---SDN, SDR, and virtualization---are expected to converge under the umbrella of 5G-SDWN for the 5G wireless networks to cater increasing demands for diverse services. Offered by these three pillars and new trends of physical layer technologies (i.e., full-duplex transmission, millimeter-wave transceivers, machine-type connections, massive multiple input multiple output (MIMO) scenarios), this new paradigm, called 5G-SDWN, provides numerous advantages, ranging from higher spectrum/energy efficiency and lower end-to-end transmission delay, to lower costs and time required for launching new applications and services \cite{7039225,7108393,7084578}.
	
Such advantages of immigrating to 5G-SDWN stem from the fact that wireless procedures and functional units of infrastructure entities can be moved into software with the aid of SDR and SDN, for both edge APs and core nodes in this generation. Furthermore, the separation between control and date planes, enabled by SDN and SDR, provides the ground to offer network and wireless virtualization \cite{7045398,7039225,6845049,7116189,costa2015software}. Allowing abstraction of resources, virtualization is a technique to share network infrastructure among different service providers (SPs) and to bound resources for a specific set of users over the concept of slicing \cite{wen2013wireless,wen2014multi,6887287,6117098}.

5G-SDWN stands at the shoulders of these three networking layers, where it inherits all their flexibilities initiated by transitioning from hardware to software-based implementation. SDR acts as a physical layer of 5G-SDWN where all the APs are reprogrammable and adjustable. SDN takes care of all management and controlling messages among nodes and cellular network functionalities. More specifically, it is a translator of all protocols, standards and vendors together in such a way that \textit{communications}, \textit{transactions} and \textit{transmissions} among different entities are technology and vendor-agnostic \cite{6461195,6819788}. Over these two layers, virtualization is surfing as an application for SPs and slices aiming to improve infrastructure utilization of 5G-SDWN.

This simple high-level structure of 5G-SDWN opens the door of cross-layer, dynamic and efficient implementation of functionalities and procedures related to each network entity from the core to the edge, while introducing the software intelligence over entire 5G-SDWN. From this programmable structure of 5G-SDWN, the centralized and comprehensive view of all network enteritis are provided, where all network infrastructure can be re-arranged and adjusted for each requested service to complete the transmission path. In other words, the flexible structure and cross-domain integrity of 5G-SDWN provide the capability to abstract resources from the infrastructure level and to deploy comprehensive resource management, in which all transmission parameters and connections are actively manipulated based on requested services, user conditions, and 5G-SDWN available resources.  Such CML resource management can handle, harmonize, and distribute user traffics among APs and core entities in such an efficient manner to cope with the underutilization of resources, QoS provisioning, and user-association from the core to the edge \cite{leon2003virtual,lin2014enabling,kang2013savi}.

To fully realize features and potentials of CML resource management in 5G-SDWN, we provide a framework to classify possible management opportunities into two categories: \textit{transmission plane management} and \textit{control plane management}. On one hand, \textit{transmission plane management} concerns with adjusting data transmission parameters including frequency (sub-carrier), time, code, antenna, power, back-haul and front-haul link, cloud, storage and application servers, gateways, and programmable switches. On the other hand, \textit{control plane management} includes data control parameters and 5G-SDWN functionalities related to each user such as  MAC allocation, QoS allocation, IP protocol allocation, security control allocation, type of virtualization and depth of slicing. In this regard, we propose a functional model for CML resource management which explains what are the required functional components and how they interact with each other.

We believe that CML resource management enabled by 5G-SDWN can considerably increase the network performance.  To manifest the importance of CML resource management, we present two case studies for association control and resource allocation in homogeneous virtualized macro-cells and small-cells (i.e., 802.11 WLANs) networks. It is shown that association control leveraging 5G-SDWN principles can be of benefit to improve overall throughput, isolation among SPs, and coverage in both types of wireless networks.

\section{Technical Review}
\begin{figure*}[!t]
	\centering
	\includegraphics[width=0.5\linewidth]{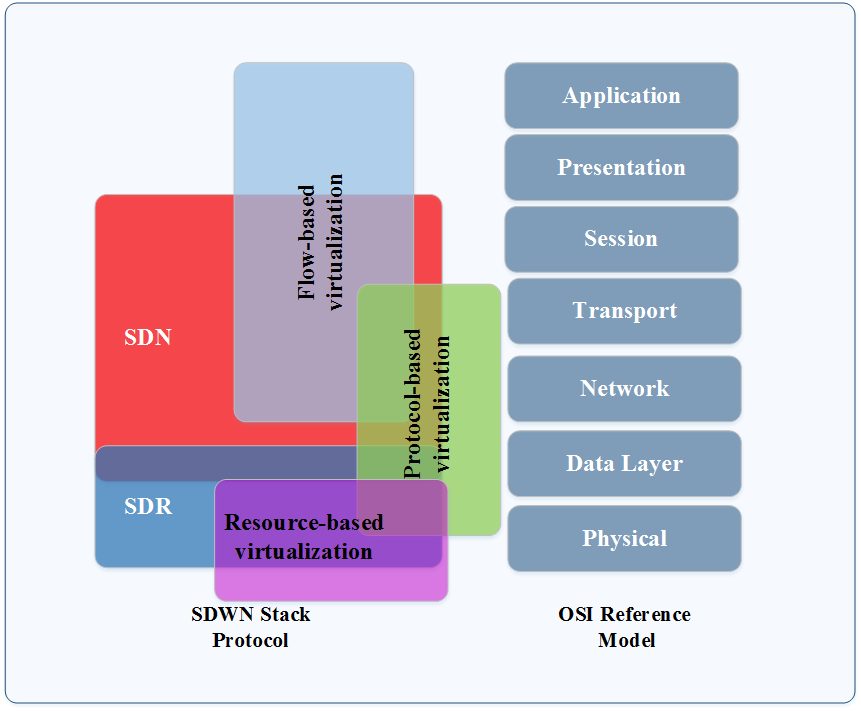}
	\caption{Stack protocol of 5G-SDWN compared to OSI model.}
\label{stack}
\end{figure*}

We believe that flexibility and software-based features of 5G-SDWN lead to more intertwined and cross-layer stack protocol design. Here, we propose a stack protocol design (which is illustrated in Figure \ref{stack}) for 5G-SDWN in comparison with the open systems interconnection (OSI) layering model. Clearly, providing \textit{software intelligence} to the physical entities of wireless networks, tasks interrelated to different OSI layers can be handled by only one layer over the stack protocol of 5G-SDWN, which can increase the processing speed and decrease the latency over 5G-SDWN-based network, leading to lower delay for requested services.    
In this section, first, we briefly review the concepts of SDN, SDR, and virtualization. As a combination of these three concepts, we present and elaborate a general architecture for 5G-SDWN and its cross-layer stack protocol. 

\subsection{SDN}\label{Sec:SDN}
Three functionality planes can be considered for a data communication network: data, control, and management planes. The data plane is responsible for (efficiently) forwarding data in network nodes, while the control plane represents the protocols, used to provide connection among data plane elements. In parallel to these two planes, the management plane monitors and configures the control functionality of elements at all layers of the protocol stack. Traditional data networks follow a tightly coupled structure of data and control planes, embedded in each network element. While supporting network's resilience, this decentralized structure causes a bottleneck for extending and updating the data networks due to its complex and relatively static architecture.

The networking paradigm, SDN, cuts this coupling design of control and data planes by \cite{7039225}: 
\begin{itemize}
	\item Removing control functionality from network nodes and turning them to simple data/packets forwarding nodes;
	\item Replacing flow-based forwarding of data decisions/routings instead of destination-based ones;
	\item Moving control logic to an external entity, called SDN controller.  
\end{itemize} 

These SDN controllers have an overall view of the network nodes. Thus, they are able to control all the nodes and their interfaces in a vendor-independent manner. This feature can be developed using high-level languages. The SDN controller translates these programs into actions for each network element and hides different interface commands for nodes. In this structure, SDN controllers have two main interfaces, called northbound and southbound. The southbound interfaces take care of the interactions between the controller, the network elements and the programmable interfaces at the edge elements. The name \textit{southbound} refers to the direction from the controller to the switches, in contrast to \textit{northbound}, which is the direction from the controller to high-level applications. The northbound interface is the
communication bridge between the controller and the control applications. The decoupling feature of SDN has shown to be fruitful in 
\begin{itemize}
	\item Streamlining implementation of network devices based on simplified stack protocols;
	\item Reducing reconfiguration, replanning, and optimization time;
	\item Facilitating deployment of new services, protocols and applications;
	\item Unifying the service platforms;
\end{itemize}

Such advantages bring more flexibility and efficiency into the SDN's than the traditional ones. Furthermore, indirectly, SDN can reduce capital and operational expenses (CAPEX and OPEX) of communications networks, promoting wireless operators to provide new services with lower prices. 

\subsection{SDR}
Delivering similar benefits as in SDN, the decoupling of the control and transmission layers can be applied in edge wireless transceivers with the help of SDR. In particular, SDR is the edge transceiver with two basic units. First, the radio unit which is responsible for transmit/receive RF signals. This unit consists of an antenna, a multi-band radio frequency (RF) module, a broadband analog to digital (A/D), and digital to analog (D/A) converters. Ideally, this unit should work at different frequencies, modes, and standards. Second, the processing unit which is responsible for all radio's operating functions including modulation/demodulation, coding/decoding, encryption/decryption, and MAC procedures. All these functionalities are implemented over programmable processing technologies such as FPGA and a generic CPU \cite{7113226}. Therefore, SDR allows traditionally hardware-integrated wireless functionalities to be controllable through software-based controllers. 
 	
There are two main architectures proposed for SDR: \textit{modal SDR} and \textit{reconfigurable SDR} \cite{7039225}. In \textit{modal SDR}, multiple implementations are integrated in a specific hardware, which can be alternated on demand. Modal SDR lacks enough flexibility as the number of required transmission standards increases. As an alternative solution, in \textit{reconfigurable SDR}, programmable hardware is employed to perform signal processing, which can be properly configured for different transmission techniques.

From the 5G-SDWN prescriptive, the SDR transceiver is a flexible and smart entity, which can enable self-organizing networking (SON) solutions and provide the portfolio of wireless resources for optimizing the network performance considering the wireless channel conditions, interference level, and QoS requirement of each user. Most importantly, it can adjust the multi-user access techniques and MAC protocols based on all mentioned parameters. Flexible reconfiguration of MAC protocol and transmission mode selection is a key enabler to reach higher spectrum and power efficiency based on the comprehensive resource management for 5G-SDWN.

For instance, different MAC protocols such as carrier sense multiple access with collision avoidance (CSMA /CA), time-division multiple access (TDMA), frequency-division multiple access (FDMA), and code-division multiple access (CDMA), orthogonal frequency-division multiple access (OFDMA) have their own pros and cons, where each works efficiently for specific network conditions. For example, TDMA suffers from under-utilization of channels when the number of users in the network is small, while CSMA/CA offers more opportunistic access which makes it more flexible for this case. In the crowded network scenario, TDMA reaches to the highest utilization, while CSMA struggles with a high collision probability. If an AP can switch between these two MAC protocols, it can effectively increase the network throughput \cite{Atoosa2015}. Also, depending on the coverage or capacity requirement, the access node can switch between an interference-limited MAC (e.g., CDMA) and a power-limited one (e.g., OFDMA or FDMA). Providing such capability to allocate the preferable MAC protocol based on the network conditions opens new doors for 5G-SDWN design and resource management, which will be discussed further in forthcoming sections. In addition, its offered flexibility inherently decreases costs of implementation, maintenance, in addition to CAPEX and OPEX in wireless networks. 

\subsection{Virtualization: Application surfing over SDN and SDR}
Virtualization allows the flexible reuse and sharing of the existing infrastructure among different SPs (also called tenants), which is another enabler for reaching higher spectrum and energy efficiency. This concept was initiated in computing and backbone networking domains, and now by cooperation of SDN and SDR, it has been extended to the wireless edge of networks. Therefore, there exists a broad range of sharing or virtualization from high-level network management, service allocation or application sharing to low-level hardware or physical resource sharing \cite{wen2013wireless,wen2014multi}.

\begin{figure*}[!t]
	\centering
	\includegraphics[width=.75\linewidth]{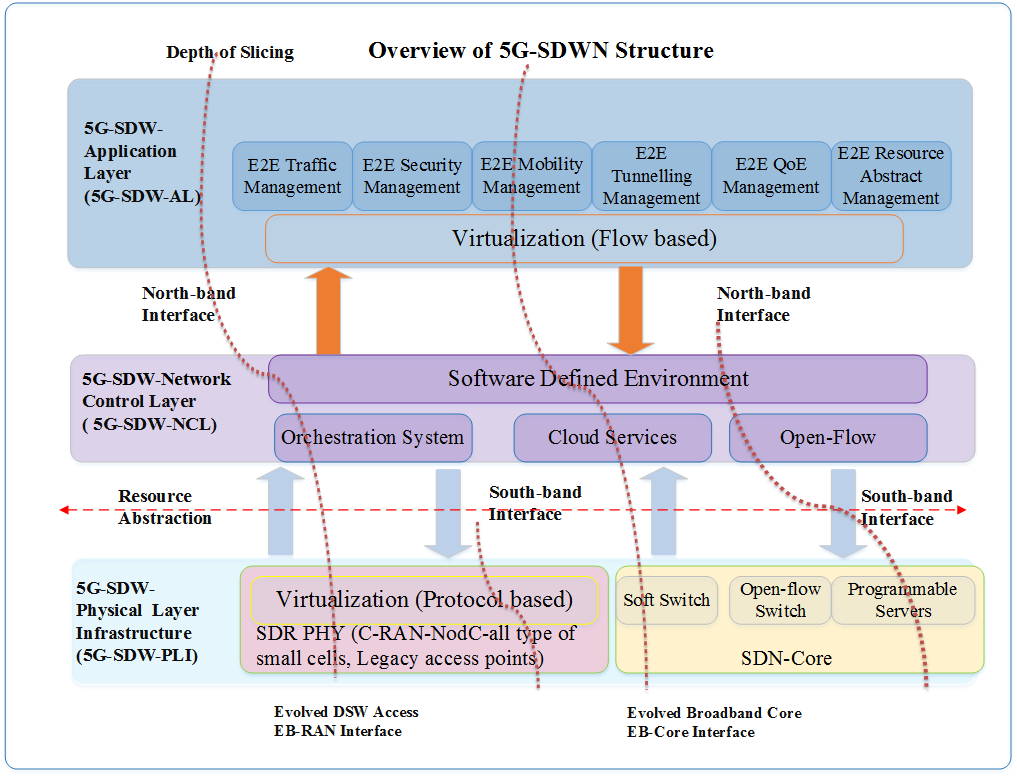}
	\caption{Illustration of 5G-SDWN generic architecture}
\label{5G-SDWN2}
\end{figure*}

In this paper, virtualization is defined as the abstraction and bundling all kinds of resources and equipment by tenants, which is considered as a main application, while networking shares context over the entire infrastructure of 5G-SDWN. In this context, the virtual instance of a set of bundled resources for one tenant is called a slice. Depending on how and in which layer resources are sliced for one tenant, the depth of slice can be determined over the stack protocol of 5G-SDWN. For example, if the SDN controller is virtualized between two tenants, the depth of slice is from the third to the second layer of 5G-SDWN stack protocol depicted in Figure \ref{stack}. The important implementation issue of slices is to provide the isolation among slices. It means that any change in one slice because of new users' arrival, mobility, and channel fluctuations, should not affect services offered to other slices. Based on the depth of slices, the virtualization for wireless networks can be categorized as follows \cite{wen2013wireless}: 
\begin{itemize}
	\item \textit{Flow-based virtualization}: This type (which is inspired by flow-based SDN technologies such as OpenFlow \cite{mckeown2008openflow}) focuses on providing isolation, scheduling, management and service differentiation between both uplink and downlink data flows from different slices. Based on this definition, this type of virtualization is developed over the SDN layer in Figure \ref{stack} or equivalently, the network layer of OSI.
	\item \textit{Protocol-based virtualization}: This category is about the isolation, customization and management of multiple wireless protocols on the same radio hardware, executed over both SDR and SDN layers of 5G-SDWN as depicted in Figure \ref{stack}. The equivalent layers of OSI are from transport layer to the physical layer. Therefore, adapting this type of virtualization requires much adjustment in the network.  
	\item \textit{Resource-based virtualization}: This model focuses on the sharing of RF front-end resources and spectrum of devices in the network, which is build based on the application of SDR and cognitive radio in wireless networks. This type of virtualization is mainly developed over the SDR and correspondingly the physical layer of OSI. 
\end{itemize}
The flexibility of SDN and SDR provided by software programmability can help to realize any combination of mentioned categories of virtualization for any tenant over the stack protocol of 5G-SDWN as it is clear from our proposed structure in Figure \ref{5G-SDWN2}. Here, the important issues are how to provide the isolation among slices, while creating more flexible and sustainable services via an efficient resource allocation and how this flexible structure can help to reach the best QoS experience from the end users' point of view. To attain these two conflicting goals, smart CML resource management over wireless networks is essential which is disused in Section III.

\begin{figure*}[!t]
	\centering
	\includegraphics[width=.85\linewidth]{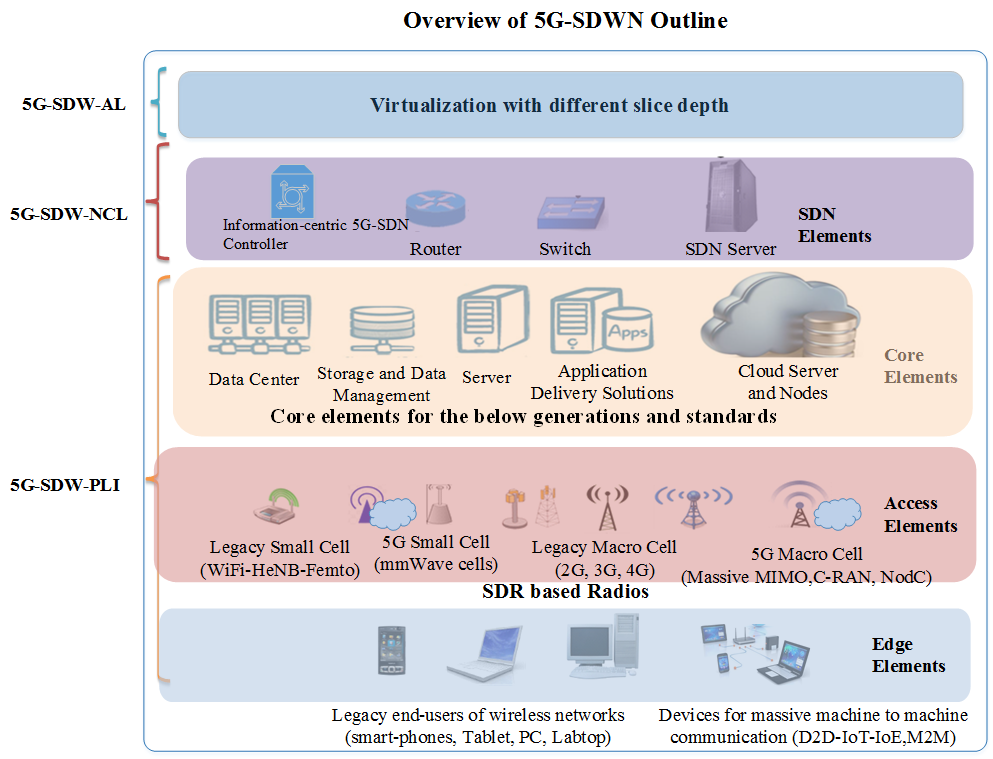}
	\caption{5G-SDWN generic Outline based on proposed architecture}
	\label{5G-SDWN3}
\end{figure*}


\section{5G-SDWN Generic Architecture }

All of the aforementioned technical advancements can be gathered under the umbrella of the 5G-SDWN which could be a major revolution of the wireless generations. A basic architecture and outline of generic 5G-SDWN structure have been put forward as shown in Figures \ref{5G-SDWN2} and \ref{5G-SDWN3}, respectively. This architecture has three main layers: 5G-SDW application layer (5G-SDW-AL), 5G-SDW network control layer (5G-SDW-NCL), and 5G-SDW physical layer infrastructure (5G-SDW-PLI).

\subsection {5G-SDW-AL} This layer contains all applications related to the wireless networks such as mobility management, connection control, and security for both access and core. In the following, the detailed functionalities, procedures, and processes related to each and every module in 5G-SDW-AL are listed.
\begin{itemize}
	\item \textit{Security management}: User authentication, encryption, and key management
	\item \textit{Mobility management}: Handover over the network from access into gateways, interactions with other 5G-SDWNs, roaming, tracking, and location update of users in different states
	\item \textit{Traffic management}: Connection management, load balancing, traffic sharing, and role control of all entities in both access and core
	\item \textit{Data tunneling management}: Routing and tunneling of data, session, or flow
	\item \textit{Quality-of-Experience (QoE) management}: QoS assignment, flow assignment, and service admission policy
	\item \textit{Resource abstract management}: MAC allocation, software selection, standard selection, and admission control.
	
\end{itemize}	
  
All applications of 5G-SDW-AL are related to the end-to-end transmission of 5G-SDWN containing both core and access. They are all defined based on the requirement of each user considering its requested services over specific virtualization scenario. The flow-based virtualization can be initiated from this layer by 5G-SDWN as demonstrated in Figure \ref{5G-SDWN3}. 

\subsection{5G-SDW-NCL} This layer contains all components and functionalities similar to the SDN controller as mentioned in Section \ref{Sec:SDN}. In this layer, an overall view of the network is achieved. Through this layer, every action is first transformed into a vendor-independent programming language and then transferred to 5G-SDW-AL (via northband interface) or to 5G-SDW-PLI (via southband interface). The abstraction of different resources in 5G-SDWN are provided here from 5G-SDW-PLI to 5G-SDW-AL. This intelligence and decoupling come from SDN (OpenFlow) and orchestration system (i.e., a broker between the applications and the network elements), and recent cloud-based protocols. All the functionalities related to the control and management of 5G-SDWN belong to this layer, including SDN server, SDN controller, routers, and switches as demonstrated in Figure \ref{5G-SDWN3}.

\subsection{5G-SDW-PLI} This layer encompasses all the physical entities required to handle all functionalities of 5G-SDWN including two major parts: 
	\begin{itemize}
	\item \textit{5G-SDW-P Access} which contains access points from different technologies and generations \cite{7113226}, e.g., legacy small cells (femto, pico, Home eNodeB (HeNB)), 5G small cells (including millimeter wave (mm-Wave) small cells), macro cells of legacy networks (e.g., BTS of 2G, NodeB (NB) of 3G, eNodeB (eNB) of LTE), macro-cells belonged to the 5G including cloud radio access network (C-RAN), NodeC, massive MIMO access points, relays, and back-haul and front-haul links \cite{7064897}. All these elements are demonstrated in the related layer in Figure \ref{5G-SDWN3}.
	\item \textit{5G-SDW-P Core} which contains all front-haul, soft switches, open-flow switches, gateways, and any programmable servers for deployment applications including home-location registration, cloud servers, and data centers, depicted in in Figure \ref{5G-SDWN3}. 
	\end{itemize}
 
To reach the seamless connection from the end-users point of view, 5G-SDWN will deal with different wireless generations, technologies, and standards specifically in 5G-SDW-P Access due to the broad range of radio access technologies. Since 5G-SDW-P Access plays a crucial role for the end-users and managing the other functionalities of 5G-SDWNs, we propose the following classification for the 5G-SDWN.
\begin{itemize}
\item  \textit{Coordinated multi-AP 5G-SDWN (CM-5G-SDWN)}: In CM-5G-SDWN, SPs within one specific region are served by the transceivers belonged to only one generation of wireless technologies, e.g., (C-RAN, massive-MIMO-based AP, 2G, 3G, LTE, or WiFi). In this setup, both coverage and capacity will be provided by the same technology for all users of networks.
\item \textit{Heterogeneous multi-tier 5G-SDWN (Het-5G-SDWN)}: In Het-5G-SDWN, different generations and technologies support the coverage and capacity of wireless access. For example, in a wireless access network, one of the legacy generations (3G or 4G) and/or C-RAN, NodeC or massive-MIMO-based APs can be deployed to provide the best coverage. This layer can be considered as a \textit{coverage/overlay layer}. Simultaneously, the high traffic hot-spots can be served by 5G small cells with LTE, WiFi and/or up-coming 5G dense deployment of small cells, as a \textit{capacity/underlay layer}.
\end{itemize}
Obviously, implementation, planning, and optimization of CM-5G-SDWN are easier than those for Het-5G-SDWN. Nevertheless, the latter delivers enormous capacity and coverage. 

In Figure \ref{5G-SDWN3}, we propose two groups of end-users for 5G-SDWN: 1) Legacy users of wireless networks, including all existing data-hungry users such as smart phones, tablets, PCs, and laptops. This type of users, which are capable to run a wide range of data applications, should connect directly to the APs of 5G-SDW-PLI. 2)  Devices for machine-type communications, including all device-to-device (D2D), machine-to-machine (M2M), Internet of things (IoT)-capable devices, which can connect to each other in addition to the 5G-SDW-PLI APs. Via CML resource management over 5G-SDWN, transmission modes and types of the connection among the latter devices can be controlled aiming to increase the spectral and energy efficiency.

According to the software-based and programmable structure of 5G-SDWN, a portfolio of network resources is available, which is abstracted from 5G-SDW-AL to the higher layers. Therefore, 5G-SDWN can leverage a CML resource management to optimize the network performance. Such integrated management of converged resources can provide energy-aware and efficient forwarding and resource allocation \cite{lin2014enabling}. For each user, such allocation will be applied over the management parameters of \textit{transmission and control planes} based on its service requirements, depth of its corresponding slice, and virtualization type. We call this procedure as ``\textit{CML resource management over 5G-SDWN}''.
	
Such resource management has been the aim of many cross-layer designs which consider resource allocation over different interfaces of traditional wireless and wired networks. Via 5G-SDWN, such cross-layer resource allocation can be developed not only over different layers of one infrastructure entity but also for those of different ones. 
Here, the main questions are \textit{whether}, \textit{how much}, and \textit{in what sense} end-users and SPs can benefit from such network-wide resource management. In the next section, we try to answer these questions by presenting two case studies on association control enabled by 5G-SDWN.

\section{Toward a converged multi-layer resource management over 5G-SDWN}

The flexible structure and cross-domain integrity of 5G-SDWN provide the capability to abstract resources from the infrastructure level and to deploy the converged multi-layer resource management over the network. In the 5G-SDWN structure, all network resources can be divided into three categories: 1) wireless resources: spectrum, transmit power, antenna, beam, time, and code; 2) computing resources including all the storage, computing units of clouds, and base-band units of C-RAN, 3) infrastructure resources, all APs, switches, links, front-haul, and back-hauls links. From the 5G-SDWN point of view, all these resources are abstracted to diffident grids and tables, which can be assigned by the CML resource management and divided to the different layers of networks. 

\subsection{Hierarchal Functional Model for CML Resource Management}

\begin{figure*}[!t]
	\centering
	\includegraphics[width=.75\linewidth]{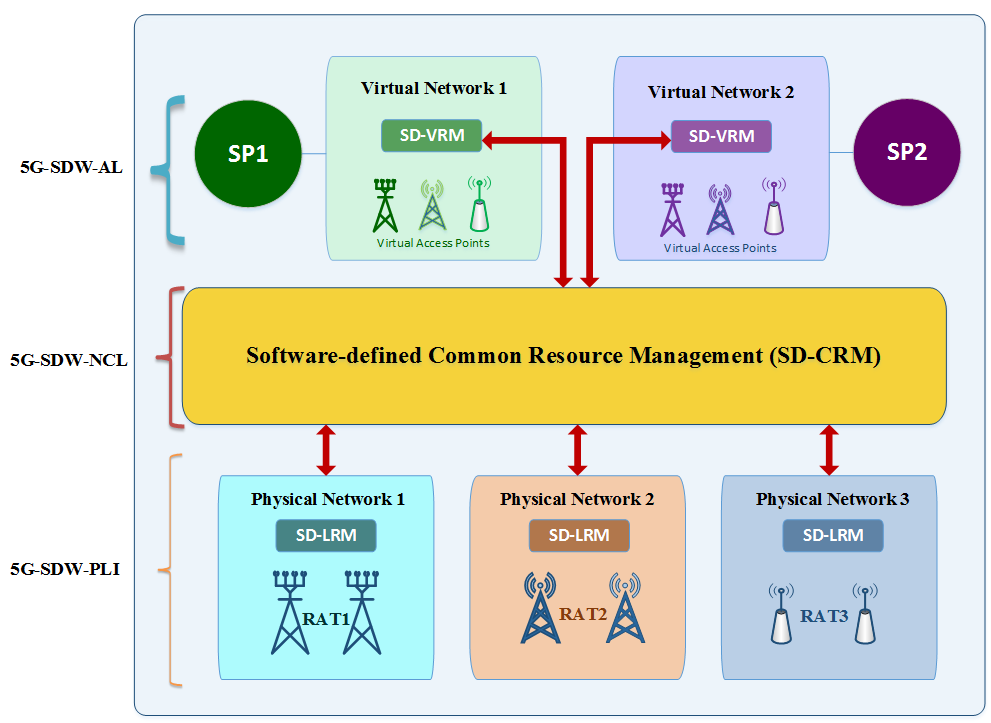}
	\caption{A functional model for CML resource management in 5G-SDWN}
	\label{Functional_Model}
\end{figure*}

Here, we present a hierarchal functional model for CML resource management in the 5G SDWN, which is illustrated in Figure \ref{Functional_Model}. 
This model consists of different elements including Software-defined Virtual Resource Management (SD-VRM), Software-defined Common Resource Management (SD-CRM), and Software-defined Local Resource Management (SD-LRM) entities. With this layering structure, CML resource management can consider each device as a packet forwarding node, where all of its transmission characteristic can be adjusted based on its grid of resources. Consequentially, the final performance of node can be translated based on the requirement of each layer as well. Via this approach, CML resource management is attainable over 5G-SDWN.  

In the following, we elaborate on the functionalities of each and every component of our proposed functional model.

\subsubsection{The SD-VRM entity} This entity is placed on top of this hierarchy and is run for each separate virtual network, which belongs to an individual SP. SD-VRM needs to translate QoS requirements and service level agreements (SLAs) for the lower levels. Depending on the level of isolation a SP is needed, scheduling end-users within the virtual network can be of its responsibility. But, such scheduling at SD-VRM is not dealing with physical resources and only virtual resources are created and allocated to provide the capacity required by the SP. The physical resource scheduling would be delegated to the lower levels.

\subsubsection{The SD-CRM entity} This entity is responsible to manage a pool of different resources, including wireless and computing resources as well as infrastructure resources of different radio access networks (RANs), in a coordinated manner.
In this layer, the physical resources of different RATs and core network are abstracted as resource blocks and SD-CRM is in charge of scheduling and resource block allocation for SPs and/or end-users. With such centralized management, it can be ensured that the resource block scheduling takes into account the QoS requirements of all slices/SPs as well as the resource availability in all SD-LRM entities. The concept of Common \textit{Radio} Resource Management (CRRM) has been already introduced by 3GPP (Third Generation Partnership Project) for heterogeneous multi-tier wireless networks  \cite{perez2008radio}.

\subsubsection{The SD-LRM entity} This entity performs the management of the resources of a specific RAN. For each physical network, SD-LRM is responsible to map the scheduling of resource blocks onto physical ones. The decision on resource block scheduling is made by SD-CRM and reported back to SD-LRM. Moreover, SD-LRM is in charge of informing SD-CRM about the available capacity and measurements (e.g., channel state information) taken at the access points.

In a heterogeneous environment where several RATs coexist and multiple SPs share the infrastructure, the modularity in this design helps to handle network-wide resource management and isolation among different SPs. In comparison to the models proposed in \cite{khatibi2014modelling,caeiro2015ondemand}, this model introduces a centralized coordinated management in the SD-CRM along with several resource manager/schedulers (i.e., SD-LRM), each exclusively for a virtual network, ensuring that SP could manage scheduling in its own virtual network.

\subsection{Main Challenges for CML Resource Management}

Such hierarchal modular CML resource management over SDWN is truly accomplished if it provides flexibility in the following aspects.
\begin{itemize}
	\item Abstracting resources in the network to reach a homogenized definition 
	\item Defining an objective function (e.g., maximizing total throughput, minimizing delay, and/or minimizing power consumption)
	\item Considering network limitations
	\item Considering network dynamics, in terms of user mobility and varying traffic arrivals 
\end{itemize}

This CML resource management can handle, harmonize, and distribute the user traffic between APs and core entities. However, such CML resource management is not trivial and straightforward over SDWN. There are technical challenges as   
\begin{itemize}
	\item Diversity of control parameters, 	
	\item Complexity of framework,
	\item Feasibility of defined optimization problem,
	\item Scalability and performance trade-off.
\end{itemize}

 More specifically, since the parameters for CML resource management are diverse, the main challenge is how to translate different layer parameters and regularize them in similar dimensions. After this step, compatible mathematical formulations need to be defined in a form of optimization problems. Such problems are generally nonconvex, combinatorial, and thus computationally complex. Due to various QoS requirements over different slices, with high probability these problems generally suffer from infeasibility issues. The other issue is how much integration over CML resource management is right and enough. Unfortunately, since number of network parameters is large, considering all of them in the problem is not reasonable and comes at the cost of high computational complexity and low scalability.

One example of such CML resource management is network-wide association control in wireless networks. In a network with densely deployed APs, before a user can access the network, it needs to make a decision about which AP to associate with. In most current vendor implementations, a connection/handover is initiated by users. In particular, users choose the AP with the highest received signal-to-noise ratio (SNR) to connect with. However, the AP with the maximum SNR may not always have enough capacity or resources to occupy an additional user. Furthermore, since the user density is often uneven in the network, the Max-SNR approach can lead to an unbalanced distribution of users among APs, causing unfairness.

Thus, delegating management rights to the network operator to decide how to associate users with APs can be useful to guarantee connectivity, manage QoS, and balance the traffic load. By remote assistance of the controller, SDWN could enable network-originated association control. In the next section, we present two case studies for association control and resource allocation enabled by SDWN in homogeneous virtualized 802.11 and cellular networks. These studies manifest the challenges and importance of network-wide resource management and association control---that can be achieved in an SDWN architecture---for service customization and QoS provisioning.

\section{User Association over SDWN}
Multi-tier multi-technology ``de-cell-ization" is a inherent structure for 5G radio access, which is highly promised by the cloud-based RAN \cite{6568922,7108393}. Consequently, the user association to the appropriate access points belonging to different tiers and technologies is of high importance. Accordingly, the user association over CML resource management not only essential but also available for 5G due to the end-to-end software defined based structure. In this section, we will present our problem formulations for user association for the scenarios of CM-5G-SDWN and present how we can overcome the computational complexity. The extension of user association over Het-5G-SDWN will remain for our future works. Notably, the user association problem over 5G and traditional wireless networks has been drawn a lot of attention recently. For instance, the user association problems for the multi-cell wireless networks equipped with massive-MIMO are studied in \cite{7247514,7153519,7166320,7314981}. In the following works, we will present how the user association factor can be defined in the wireless networks, which can combine and interrelate different implementation limitations in this context with considering new sets of constraints.
 
\subsection{Association and Airtime Control in Virtualized 802.11 Networks}
In virtualized 802.11 WLANs, transmissions of different virtual WLANs (V-WLANs) are closely coupled, although administrative virtualization (i.e., one physical AP advertises multiple service set identifiers (SSIDs)) can already differentiate groups of flows. With a CSMA-based MAC, unavoidable collisions act to couple the transmissions of different V-WLANs. Moreover, since the network capacity is shared yet constrained, the increase of traffic in one V-WLAN may reduce the available network capacity to another. Thus, an efficient resource allocation among V-WLANs is essential to manage the MAC-layer couplings.

To overcome such MAC-layer couplings and balance the load, we propose a user-level management approach in virtualized 802.11 networks. We aim to generalize the association control problem by adjusting the transmission probability of each user at any AP, rather than selecting one AP to associate with. Thus, we define $\tau_i^a$ ($0\leq \tau_i^a\leq 1$) as the probability that user $i$ attempts to transmit at AP $a$ in a general time-slot. Taking into account user transmission rates and SP airtime reservations, in this approach, transmission probability of each user at each AP is optimally adjusted to maximize the overall network throughput, while keeping a total airtime guarantee for each SP. Details of the optimization problem and the developed iterative algorithm to obtain optimal transmission probabilities can be found in our study \cite{Derakhshani2014}.

Nevertheless, in 802.11 MAC protocols such as enhanced distributed channel access (EDCA), the transmission probabilities of users are not directly controllable. Instead, what we can manipulate are the MAC-layer parameters (e.g., minimum contention window and inter-frame spaces) to achieve the optimal performance. It is shown that minimum contention window is the best parameter to be controlled aiming to adjust $\tau_i^a$ to a desired value for its whole feasibility region.

\begin{figure*}
	\centering
	\subfloat[Illustration of virtualized 802.11 network]{
		\label{Fig:SysWiFi}
		\includegraphics[width=0.45\linewidth]{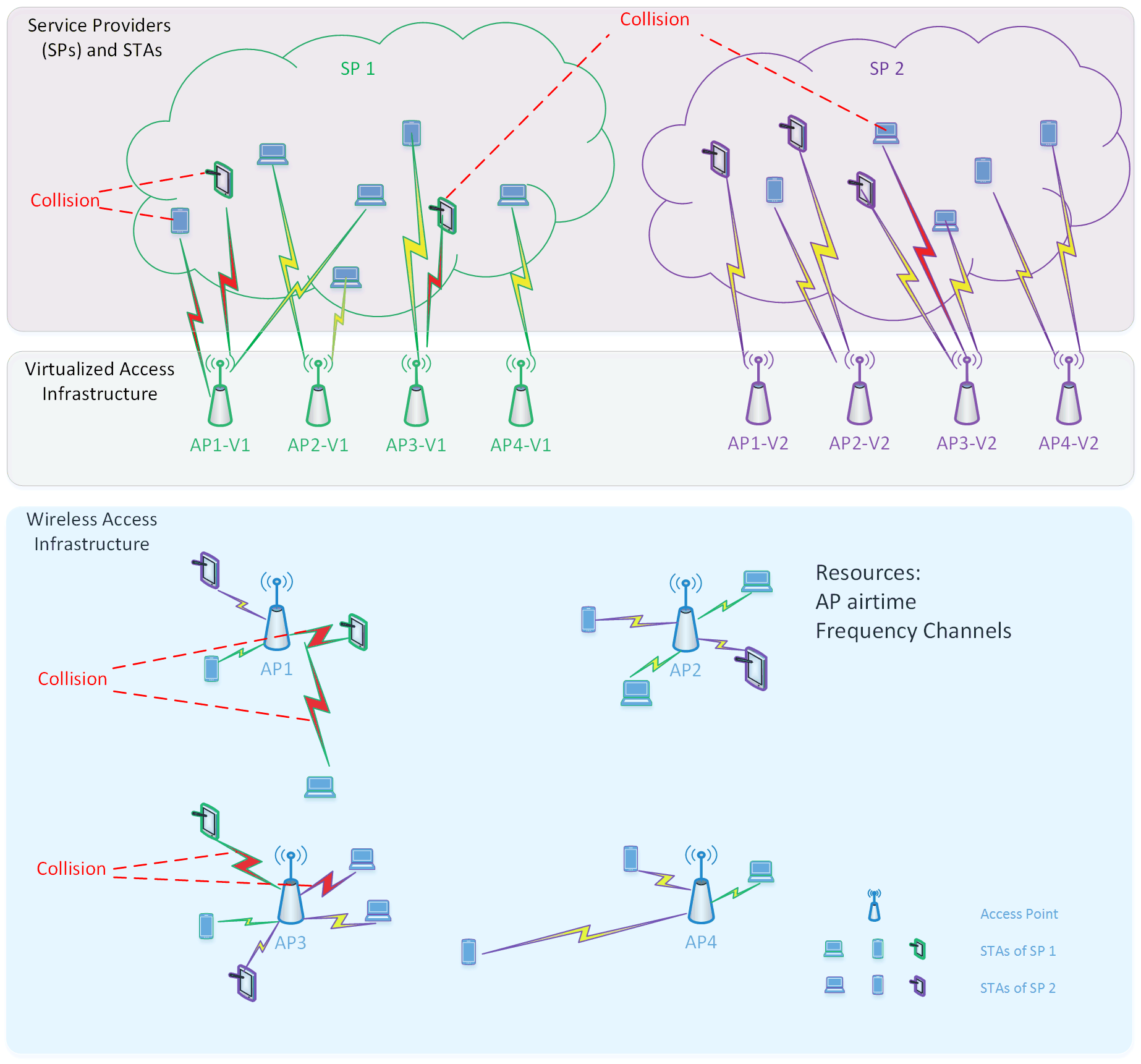}}
	\subfloat[Illustration of virtualized cellular network]{
		\label{Fig:SysCellular}
		\includegraphics[width=0.45\linewidth]{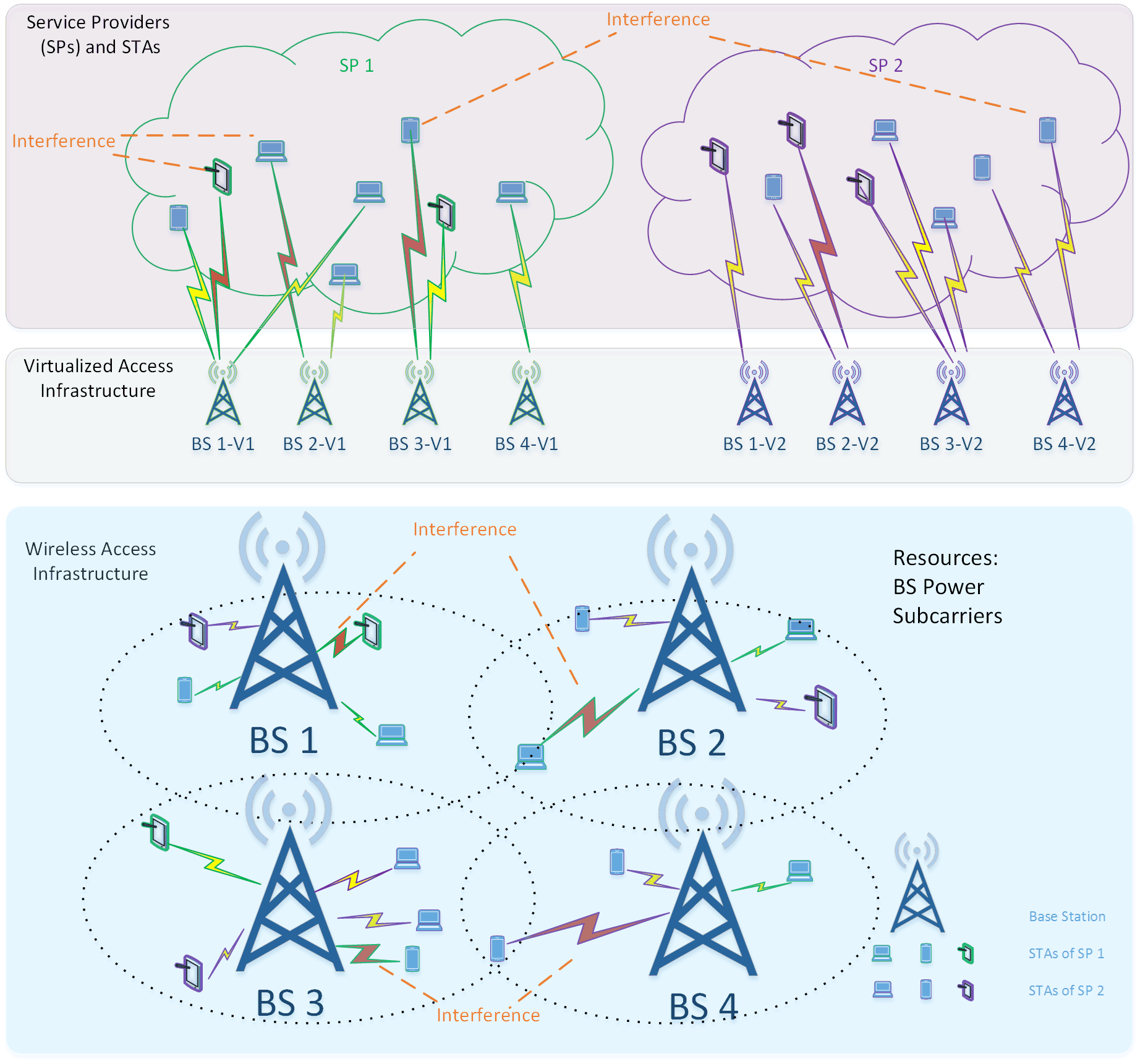}}\\
	\subfloat[Total throughput vs. user density for different $\rho_1$]{
		\label{Fig:Throughput_vs_Density_Balance}
		\includegraphics[width=0.45\linewidth]{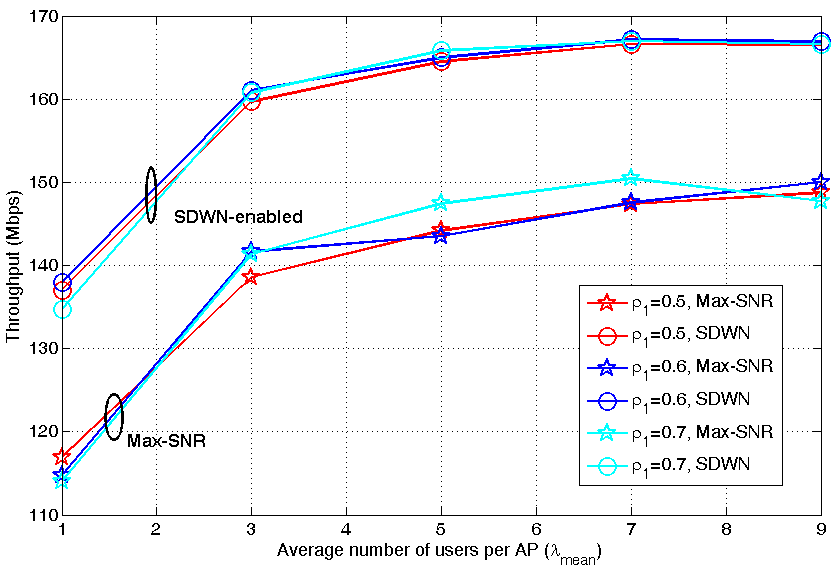}}
	\subfloat[Throughput distribution for cell-edge users]{
		\label{CDF}
		\includegraphics[width=0.45\linewidth]{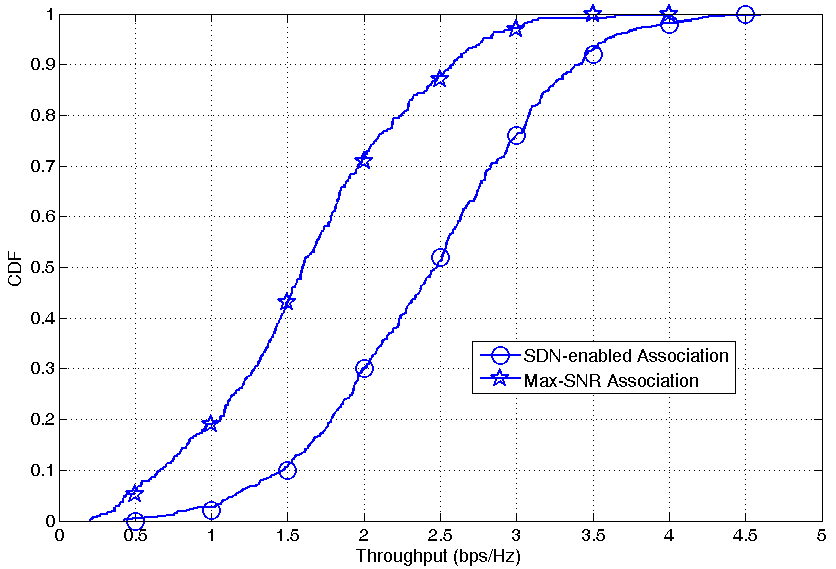}} \\
	\subfloat[Fairness vs. user density for different $\rho_1$]{
		\label{Fig:Fairness_vs_Density_Balance}
		\includegraphics[width=0.45\linewidth]{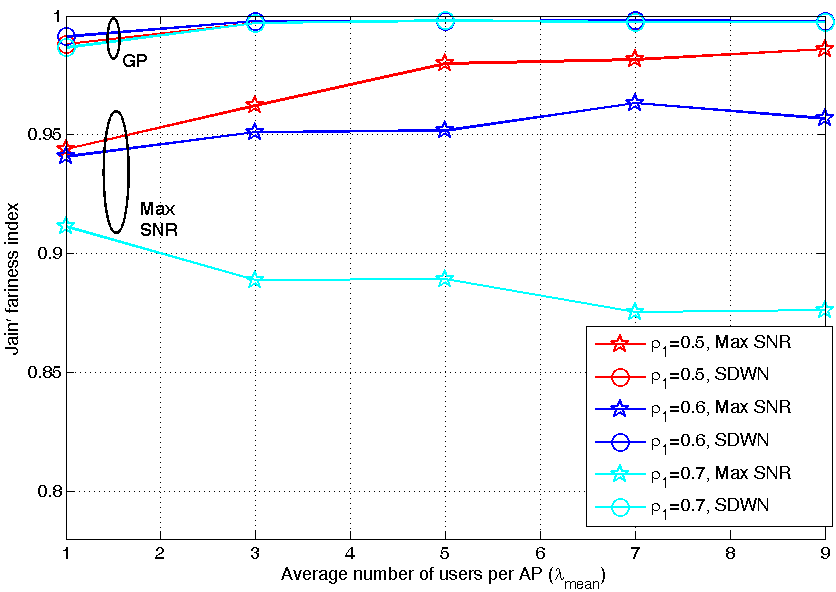}}
	\subfloat[Total throughput achieved vs. number of users]{
		\label{nonuniform_users}
		\includegraphics[width=0.45\linewidth]{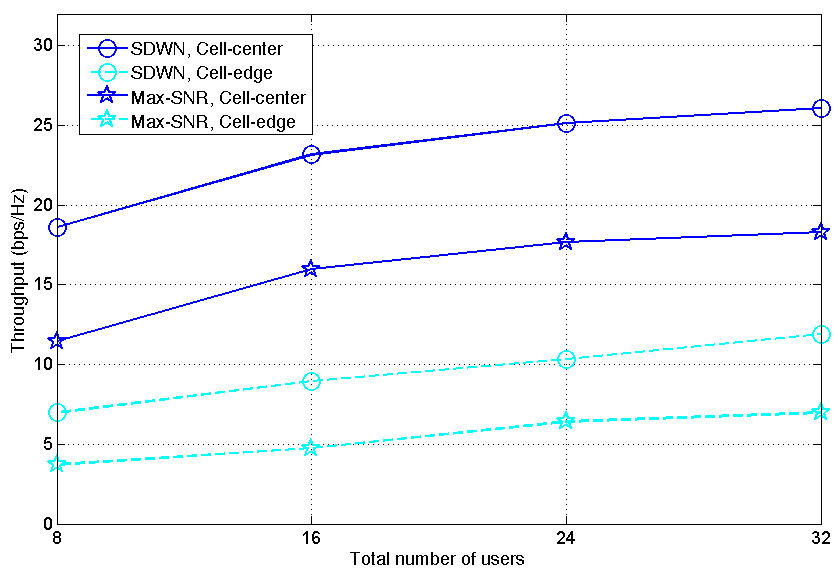}}
	\caption{System model and performance evaluation for homogeneous virtualized 802.11 and cellular networks}
	\label{case_study}
	\vspace{10mm}
\end{figure*}

Here, the performance of the proposed SDWN-enabled and conventional Max-SNR association approaches are compared in two examples under different user density and SP load. Consider a network with 4 APs as shown in Figure \ref{Fig:SysWiFi}. To eliminate interference between the transmission of different APs, four non-overlapping $20~\text{MHz}$ channels are assigned to four APs. The users are distributed in the entire area according to the two-dimensional Poisson point process (PPP). Let define $\rho_1$ (also referred to as SP 1 load) as the ratio of number of users served by SP 1 to the total number of users in the network.

The first example in Figure \ref{Fig:Throughput_vs_Density_Balance} shows the total throughput achieved by the two association algorithms versus $\lambda_{\text{mean}}$ (which represents the average number
of users per AP) for a homogeneous user distribution. For a fixed $\rho_1$, the total throughput by both algorithms increases with the user density. But, the throughput increase rate is decreasing with $\lambda_{\text{mean}}$. This is because the wireless channel is underutilized when the user density is low. Thus, the increase in the user density will improve the total throughput. But, when the user density is large, increasing the user density further will result in a higher collision probability, and hence, slow down the total throughput improvement. For any fixed $\rho_1$, it is shown that SDWN-enabled association significantly improves the total throughput compared with the Max-SNR.

The second example in Figure \ref{Fig:Fairness_vs_Density_Balance} measures the fairness by employing the Jain's fairness index over $T_k=\sum_{i\in \mathcal{S}_k,a\in \mathcal{A}}T_i^a$, which is the achieved throughput for all the users of SP $k$. It should be noted that $\mathcal{S}_k$ denotes the set of users belonging to SP $k$ and $\mathcal{A}$ represents the set of all APs in the networks. From Figure \ref{Fig:Fairness_vs_Density_Balance}, it is clear that the proposed SDWN-enabled association approach can always guarantee perfect fairness between the SPs regardless of the user density or $\rho_1$. The achieved fairness level by Max-SNR association is always worse than SDWN-enabled, especially when the user load is highly unbalanced between SPs (i.e., $\rho_1$ is not close to $0.5$).

\subsection{User Association and Resource Allocation in Virtualized Cellular Networks}
Here, we study one of the applications of SDWN in a cellular radio access network for traffic shaping at the user level. Using the SDWN and the centralized view it provides, data plane cooperation across BSs can be realized for performance optimization in a multi-cell scenario. One of such optimization is coordinated association control in virtualized cellular networks, aiming to balance the load, manage the intra-cell interference and guarantee QoS requirements customized for each SP. 

By allowing data to be controlled centrally, we present a cohesive approach on wireless resource management (power and sub-carrier allocation) and user association. Specifically, in a multi-cell OFDMA-based network, problem of joint user association and power/subcarrier assignment is optimally solved. Details of the optimization problem and the developed iterative algorithm to obtain optimal resource allocation can be found in our study \cite{Parsaeefard2015}.

Here, the performance of the proposed SDWN-enabled and conventional Max-SNR association approaches are compared in two examples. Consider a virtualized cellular network with $4$ BSs with $4$ sub-carriers and $2$ slices. 4 APs are deployed as shown in Figure \ref{Fig:SysCellular}. 

In any cellular network, the coverage is one of the most important planning parameters which can be measured by the SINR or achieved total throughput of users at the cell boundaries. To study the performance of SDWN-enabled association control to increase the coverage of our scenario, we consider the simulation setup in which majority of users’ are located in the cell-edge, consequently, these users experience high interference from other BSs.

The first example in Figure \ref{CDF} demonstrates the cumulative distribution function (CDF) of the total throughput of cell-edge users for both SDWN-enabled and Max-SNR algorithms. It can be seen that SDWN-enabled association outperforms Max-SNR for the cell-edge users where 50\% of users in the cell-edge achieve a throughput of $2.5$ bps/Hz in the case of SDWN-enabled association, while their throughput is around $1.5$ bps/Hz in the case of Max-SNR. However, the performance of both algorithms are similar for the cell-center users. It is because via user-association in SDWN-enabled association, the interference among different cell can be controlled while Max-SNR cannot control the interference through the connectivity of users to different BS and it is per-determined by the received SNR of reference signal. In other words, SDWN-enabled association can provide the better coverage even for cell-edge users for virtualized multi-cellular networks which is very desirable from implementation perspective.

The second example in Figure \ref{nonuniform_users} investigates the total achieved throughput with respect to the number of users at the cell edge. Obviously, SDWN-enabled association can consistently improve the performance of cell-edge users and at the same time maintain desirable throughput of overall networks regardless of the user deployment density compared to the Max-SNR.

\section{Conclusion and Future Works}
In this paper, the simple, efficient and integrated structure of 5G-SDWN has been studied based on its three pillars, SDN, SDR, and virtualization. For this networking paradigm, we have provided a general architecture and discussed that the 5G-SDWN structure enables converged multi-layer resource management. Via two case-studies, we have highlighted advantages of such resource management for 5G-SDWN. In particular, it is shown how it can increase the coverage and capacity while providing the requested QoS.

We believe that the CML resource management over 5G-SDWN is at the first stage of development, encountering numerous issues and encouraging for future research. From the soft, integrated, centralized and cross-layer structure of 5G-SDWN, it is expected that CML resource management cannot be an evolutionary version of traditional problems in wireless networks, while it needs more revolutionary movement and thoughts. 

In addition to the discussed issues, another implementation challenge is deriving all the required information for each user. Clearly, network can hardly obtain the perfect and complete information. Therefore, robust and learning approaches are required to consider the uncertainty on the system parameters and to derive all individual features of users or system information. Solving this type of general optimization problems necessitates more sophisticated mathematical tools and programming algorithms. Last but not least, cooperation and connection of 5G-SDWN with traditional wireless networks cause new challenges.

In the future, we shall focus on the performance evaluation
of 5G-SDWN considering other parameters from transmission and control planes. In addition, we shall study possible mathematical tools and approaches to solve such CML resource management over 5G-SDWN. The combination of new radio transmission concepts over access of 5G-SDWN, such as full-duplex, milliliter Wave, massive MIMO, and device-to-device communications, will be investigated as the future extension of proposed two case studies.


\begin{thebibliography}{10}
	\providecommand{\url}[1]{#1}
	\csname url@samestyle\endcsname
	\providecommand{\newblock}{\relax}
	\providecommand{\bibinfo}[2]{#2}
	\providecommand{\BIBentrySTDinterwordspacing}{\spaceskip=0pt\relax}
	\providecommand{\BIBentryALTinterwordstretchfactor}{4}
	\providecommand{\BIBentryALTinterwordspacing}{\spaceskip=\fontdimen2\font plus
		\BIBentryALTinterwordstretchfactor\fontdimen3\font minus
		\fontdimen4\font\relax}
	\providecommand{\BIBforeignlanguage}[2]{{%
			\expandafter\ifx\csname l@#1\endcsname\relax
			\typeout{** WARNING: IEEEtran.bst: No hyphenation pattern has been}%
			\typeout{** loaded for the language `#1'. Using the pattern for}%
			\typeout{** the default language instead.}%
			\else
			\language=\csname l@#1\endcsname
			\fi
			#2}}
	\providecommand{\BIBdecl}{\relax}
	\BIBdecl
	
	\bibitem{7113228}
	C.~Liang, F.~Yu, and X.~Zhang, ``Information-centric network function
	virtualization over {5G} mobile wireless networks,'' \emph{IEEE Network},
	vol.~29, no.~3, pp. 68--74, May 2015.
	
	\bibitem{7108393}
	E.~Hossain and M.~Hasan, ``5g cellular: key enabling technologies and research
	challenges,'' \emph{IEEE Instrumentation Measurement Magazine}, vol.~18,
	no.~3, pp. 11--21, June 2015.
	
	\bibitem{7113226}
	S.~Sun, M.~Kadoch, L.~Gong, and B.~Rong, ``Integrating network function
	virtualization with {SDR} and {SDN} for {4G/5G} networks,'' \emph{IEEE
		Network}, vol.~29, no.~3, pp. 54--59, May 2015.
	
	\bibitem{7120046}
	N.~Zhang, N.~Cheng, A.~Gamage, K.~Zhang, J.~Mark, and X.~Shen, ``Cloud assisted
	{HetNets} toward {5G} wireless networks,'' \emph{IEEE Communications
		Magazine}, vol.~53, no.~6, pp. 59--65, June 2015.
	
	\bibitem{peng2015system}
	M.~Peng, Y.~Li, Z.~Zhao, and C.~Wang, ``System architecture and key
	technologies for {5G} heterogeneous cloud radio access networks,'' \emph{IEEE
		Network}, vol.~29, no.~2, pp. 6--14, 2015.
	
	\bibitem{7039225}
	D.~Macedo, D.~Guedes, L.~Vieira, M.~Vieira, and M.~Nogueira, ``Programmable
	networks-from software defined radio to software defined networking,''
	\emph{{IEEE} Commun. Surveys Tuts.}, 2015.
	
	\bibitem{7084578}
	M.~Gerasimenko, D.~Moltchanov, R.~Florea, S.~Andreev, Y.~Koucheryavy,
	N.~Himayat, S.-P. Yeh, and S.~Talwar, ``Cooperative radio resource management
	in heterogeneous cloud radio access networks,'' \emph{IEEE Access}, vol.~3,
	pp. 397--406, 2015.
	
	\bibitem{7045398}
	M.~R. Sama, L.~M. Contreras, J.~Kaippallimalil, I.~Akiyoshi, H.~Qian, and
	H.~Ni, ``Software-defined control of the virtualized mobile packet core,''
	\emph{{IEEE} Commun. Mag.}, vol.~53, no.~2, pp. 107--115, Feb. 2015.
	
	\bibitem{6845049}
	C.~Bernardos, A.~De~La~Oliva, P.~Serrano, A.~Banchs, L.~Contreras, H.~Jin, and
	J.~Zúniga, ``An architecture for software defined wireless networking,''
	\emph{IEEE Wireless Commun.}, vol.~21, no.~3, pp. 52--61, Jun. 2014.
	
	\bibitem{7116189}
	C.~Niephaus, G.~Ghinea, O.~Aliu, S.~Hadzic, and M.~Kretschmer, ``{SDN} in the
	wireless context - {T}owards full programmability of wireless network
	elements,'' in \emph{IEEE Conf. Network Softwarization (NetSoft)}, April
	2015, pp. 1--6.
	
	\bibitem{costa2015software}
	J.~Costa-Requena, V.~F. Guasch, and J.~L. Santos, ``Software defined networks
	based 5g backhaul architecture,'' in \emph{Proceedings of the 9th
		International Conference on Ubiquitous Information Management and
		Communication}.\hskip 1em plus 0.5em minus 0.4em\relax ACM, 2015, p.~35.
	
	\bibitem{wen2013wireless}
	H.~Wen, P.~K. Tiwary, and T.~Le-Ngoc, \emph{Wireless Virtualization}.\hskip 1em
	plus 0.5em minus 0.4em\relax Springer, 2013.
	
	\bibitem{wen2014multi}
	------, ``Multi-perspective virtualization and software-defined infrastructure
	framework for wireless access networks,'' \emph{Mobile Networks and
		Applications}, pp. 1--13, 2014.
	
	\bibitem{6887287}
	C.~Liang and F.~Yu, ``Wireless network virtualization: A survey, some research
	issues and challenges,'' \emph{{IEEE} Commun. Surveys Tuts.}, vol.~17, no.~1,
	pp. 358--380, Firstquarter 2015.
	
	\bibitem{6117098}
	{R. Kokku \textit{et al.}}, ``{NVS}: A substrate for virtualizing wireless
	resources in cellular networks,'' \emph{{IEEE/ACM} Trans. Netw.}, vol.~20,
	no.~5, Oct. 2012.
	
	\bibitem{6461195}
	H.~Kim and N.~Feamster, ``Improving network management with software defined
	networking,'' \emph{{IEEE} Commun. Mag.}, vol.~51, no.~2, pp. 114--119, Feb.
	2013.
	
	\bibitem{6819788}
	F.~Hu, Q.~Hao, and K.~Bao, ``A survey on software-defined network and
	{OpenFlow}: From concept to implementation,'' \emph{{IEEE} Commun. Surveys
		Tuts.}, vol.~16, no.~4, pp. 2181--2206, Fourthquarter 2014.
	
	\bibitem{leon2003virtual}
	A.~Leon-Garcia and L.~G. Mason, ``Virtual network resource management for
	next-generation networks,'' \emph{IEEE Communications Magazine}, vol.~41,
	no.~7, pp. 102--109, 2003.
	
	\bibitem{lin2014enabling}
	T.~Lin, J.-M. Kang, H.~Bannazadeh, and A.~Leon-Garcia, ``Enabling {SDN}
	applications on software-defined infrastructure,'' in \emph{IEEE Network
		Operations and Management Symposium (NOMS)}, 2014, pp. 1--7.
	
	\bibitem{kang2013savi}
	J.-M. Kang, H.~Bannazadeh, and A.~Leon-Garcia, ``{SAVI} testbed: Control and
	management of converged virtual {ICT} resources,'' in \emph{IFIP/IEEE Intl.
		Symposium Integrated Network Management}, 2013, pp. 664--667.
	
	\bibitem{Atoosa2015}
	A.~Shoaei~Dalili, M.~Derakhshani, S.~Parsaeefard, and T.~Le-Ngoc,
	``Learning-based hybrid tdma-csma mac protocol for virtualized 802.11
	wlans,'' in \emph{Proc. of PIMRC}, Sep. 2015.
	
	\bibitem{mckeown2008openflow}
	N.~McKeown, T.~Anderson, H.~Balakrishnan, G.~Parulkar, L.~Peterson, J.~Rexford,
	S.~Shenker, and J.~Turner, ``Open{F}low: enabling innovation in campus
	networks,'' \emph{ACM SIGCOMM Computer Communication Review}, vol.~38, no.~2,
	pp. 69--74, 2008.
	
	\bibitem{7064897}
	M.~Peng, Y.~Li, Z.~Zhao, and C.~Wang, ``System architecture and key
	technologies for {5G} heterogeneous cloud radio access networks,'' \emph{IEEE
		Network}, vol.~29, no.~2, pp. 6--14, Mar. 2015.
	
	\bibitem{perez2008radio}
	J.~Perez-Romero, X.~Gelabert, and O.~Sallent, ``Radio resource management for
	heterogeneous wireless access networks,'' \emph{Heterogeneous Wireless Access
		Networks Architectures and Protocols, Hossain, E.(Ed.). Springer Science, New
		York}, pp. 133--166, 2008.
	
	\bibitem{khatibi2014modelling}
	S.~Khatibi and L.~M. Correia, ``Modelling of virtual radio resource management
	for cellular heterogeneous access networks,'' in \emph{Proc. of PIMRC}, Sep.
	2014.
	
	\bibitem{caeiro2015ondemand}
	L.~Caeiro, F.~D. Cardoso, and L.~M. Correia, ``Ondemand virtual radio resource
	allocation for wireless access,'' \emph{Wireless Personal Communications},
	vol.~82, no.~4, pp. 2431--2456, 2015.
	
	\bibitem{6568922}
	P.~Demestichas, A.~Georgakopoulos, D.~Karvounas, K.~Tsagkaris, V.~Stavroulaki,
	J.~Lu, C.~Xiong, and J.~Yao, ``{5G} on the horizon: Key challenges for the
	radio-access network,'' \emph{IEEE Vehicular Technology Magazine}, vol.~8,
	no.~3, pp. 47--53, Sep. 2013.
	
	\bibitem{7247514}
	A.~Gotsis, S.~Stefanatos, and A.~Alexiou, ``Optimal user association for
	massive mimo empowered ultra-dense wireless networks,'' in
	\emph{Communication Workshop (ICCW), 2015 IEEE International Conference on},
	Jun. 2015, pp. 2238--2244.
	
	\bibitem{7153519}
	D.~Liu, L.~Wang, Y.~Chen, T.~Zhang, K.~K. Chai, and M.~Elkashlan, ``Distributed
	energy efficient fair user association in massive mimo enabled hetnets,''
	\emph{IEEE Communications Letters}, vol.~19, no.~10, pp. 1770--1773, Oct.
	2015.
	
	\bibitem{7166320}
	Y.~Lim, C.~Chae, and G.~Caire, ``Performance analysis of massive {MIMO} for
	cell-boundary users,'' \emph{IEEE Transactions on Wireless Communications},
	vol.~PP, no.~99, pp. 1--1, 2015.
	
	\bibitem{7314981}
	D.~Bethanabhotla, O.~Bursalioglu, H.~Papadopoulos, and G.~Caire, ``Optimal
	user-cell association for massive {MIMO} wireless networks,'' \emph{IEEE
		Transactions on Wireless Communications}, vol.~PP, no.~99, pp. 1--1, 2015.
	
	\bibitem{Derakhshani2014}
	M.~Derakhshani, X.~Wang, T.~Le-Ngoc, and A.~Leon-Garcia, ``Virtualization of
	multi-cell 802.11 networks: Association and airtime control,'' \emph{arXiv
		preprint arXiv:1508.03554}, 2015.
	
	\bibitem{Parsaeefard2015}
	S.~Parsaeefard, R.~Dawadi, M.~Derakhshani, and T.~Le-Ngoc, ``Joint
	user-association and resource-allocation in virtualized wireless networks,''
	\emph{arXiv preprint arXiv:1508.06307}, 2015.
	
\end{thebibliography}
\end{document}